\documentclass[fleqn]{article}

\usepackage{INTERSPEECH2019}
\usepackage{cite}
\usepackage{multirow}

\title{MSR-86K: An Evolving, Multilingual Corpus with 86,300 Hours of Transcribed Audio for Speech Recognition Research}
\name{Song Li, Yongbin You, Xuezhi  Wang,  Zhengkun Tian, Ke Ding, Guanglu Wan}
\address{Meituan}
\email{\{lisong39, dingke02, wanguanglu\}@meituan.com}

\begin{document}

\maketitle

\begin{abstract}
\label{sec:intro}
Recently, multilingual artificial intelligence assistants, exemplified by ChatGPT, have gained immense popularity. As a crucial gateway to human-computer interaction, multilingual automatic speech recognition (ASR) has also garnered significant attention, as evidenced by systems like Whisper. However, the proprietary nature of the training data has impeded researchers' efforts to study multilingual ASR. This paper introduces MSR-86K, an evolving, large-scale multilingual corpus for speech recognition research. The corpus is derived from publicly accessible videos on YouTube, comprising 15 languages and a total of 86,300 hours of transcribed ASR data. We also introduce how to use the MSR-86K corpus and other open-source  corpora to train a robust multilingual ASR model that is competitive with Whisper.  MSR-86K will be publicly released on  HuggingFace\footnote{https://huggingface.co/datasets/Alex-Song/MSR-86K}, and we believe that such a large corpus will pave new avenues for research in multilingual ASR.  

\end{abstract}

\section{Introduction}
Thanks to the rapid development of deep learning, research in speech recognition has gradually shifted from hybrid systems based on Hidden Markov Models  to end-to-end ASR systems entirely built on neural networks\cite{graves2006connectionist,graves2012sequence,chan2016listen,zhang2023google,radford2023robust,pratap2023scaling}. In fact, the swift progress of end-to-end ASR has also benefited from the contribution of open-source corpora, such as the commonly used LibriSpeech\cite{panayotov2015librispeech} and GigaSpeech\cite{chen2021gigaspeech} for English, as well as  AISHELL\cite{bu2017aishell} and WenetSpeech\cite{zhang2022wenetspeech} for Chinese.  These open-source corpora have facilitated research in the field of speech recognition by both academia and industry. In the multilingual domain, the Common Voice\cite{ardila2019common} project, alongside the multilingual LibriSpeech (MLS) \cite{pratap2020mls} corpus released by Meta, has greatly promoted research in multilingual ASR. In recent times, the success of OpenAI's Whisper\cite{radford2023robust} model has demonstrated that big data combined with large models can yield improved performance. However, Whisper has not made its training data public, hindering researchers' ability to replicate the results. The MSR-86K corpus introduced in this paper aims to bridge this gap, further advancing research in multilingual ASR.

\begin{table}[htbp]
  \centering
  \caption{Compare MSR-86K to common public multilingual ASR corpora.}
\resizebox{80mm}{18mm} {
    \begin{tabular}{lccll}
    \toprule
    \textbf{Corpus} & \multicolumn{1}{l}{\textbf{\# Languages}} & \multicolumn{1}{p{7.335em}}{\textbf{~~~~Total Hours \newline\makebox[0pt][l]{\hspace{2.5mm}(Transcribed)}}} & \textbf{Domains} & \textbf{Speech Type} \\
    \midrule
    BABEL\cite{gales2014speech} & 17    & 1k    & Conversational & Spontaneous \\
    Common Voicecite\cite{ardila2019common} & 112   & 18k   & Open domain & Read \\
    MLS\cite{pratap2020mls}   & 8     & 50.5k & Audiobook & Read \\
    FLEURS & 102   & 1.4k  & Wikipedia & Read \\
    CMU Wilderness\cite{black2019cmu} & 700   & 14k   & Religion & Read \\
    CoVoST-2\cite{wang2020covost} & 22    & 2.9k  & Open domain &  Read \\
    Europarl-ST\cite{iranzo2020europarl} & 6     & 500   & Parliament & Spontaneous \\
    MuST-C\cite{cattoni2021must} & 9     & 385   & TED talks & Spontaneous \\
    mTEDx\cite{salesky2021multilingual} & 9     & 1k    & TED talks & Spontaneous \\
    VoxPopuli\cite{wang2021voxpopuli} & 16    & 1791  & Parliament & Spontaneous \\
    CVSS\cite{jia2022cvss}  & 22    & 1.1k  & Open domain & Read/Synthetic \\
    MSR-86K (ours) & 15    & 86.3k & YouTube & Spontaneous \\
    \bottomrule
    \end{tabular}}%
  \label{tab:addlabel}%
\vspace{-0.5cm}
\end{table}%

Existing multilingual ASR corpora have two main shortcomings: firstly, most corpora are dominated by English and Western European languages, lacking sufficient linguistic diversity. Secondly, although some corpora have a broad coverage of languages, the duration of recordings for each language is often minimal, insufficient for building a usable ASR system. The MSR-86K corpus addresses these issues by ensuring substantial coverage of languages and providing enough data per language to independently train a robust ASR system.  We constructed a series of protocols to automatically retrieve publicly accessible videos from YouTube and set up a data processing pipeline to automatically generate the MSR-86K corpus, significantly reducing the costs associated with data collection and labeling. Table 1 illustrates the distinctions between our MSR-86K and other public multilingual ASR corpora.

Whisper is an excellent multilingual model, but its best-performing variant has a large number of parameters, which results in slower inference speed and greater memory overhead. In this paper, we introduce how to use easily accessible unsupervised data for pre-training and fine-tuning with MSR-86K and other open-source corpora to build a robust multilingual ASR model that is faster, smaller in size, and has performance that matches or even exceeds that of the Whisper large model.

The rest of the paper is organized as follows. In Section 2, the process of constructing the MSR-86K corpus  is described. In Section 3, we introduce our experiments and discussions.  Finally, the paper is concluded in Section 4.

\section{Corpus Construction}
This section describes the major steps involved in creating the MSR-86K corpus, and Figure 1 illustrates this process.

\begin{figure*}[h]
\vspace{-2cm}
  \centering
  \includegraphics[width=1.0\linewidth]{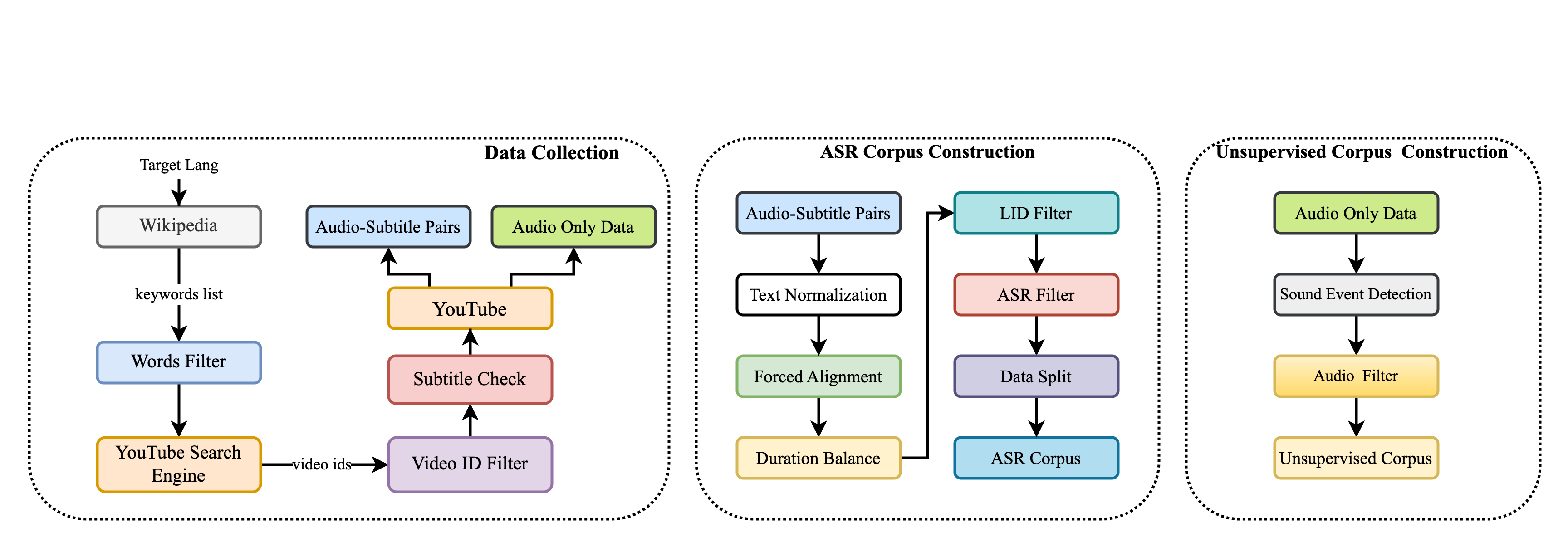}
  \caption{The construction process of the MSR-86K corpus.}
  \label{fig:speech_production}
\end{figure*}

\subsection{Data collection}
\textbf{Creating keyword lists.} First, we start by generating a preliminary list of keywords through querying Wikipedia articles in the target language. Recognizing the presence of numerous non-target language terms within these entries, we then implement a keyword filtering module to refine our list. The module selectively filters and retains terms that are likely to be significant keywords, ensuring relevance in the target language.

\textbf{Retrieving video IDs.} Next, we use the YouTube search engine to search the keyword list, obtaining a list of video IDs. Since different keywords may lead to the same videos, it is necessary to deduplicate the video ID list. We hope to share the dataset, so we further filter out videos that are available for public download, and remove private, paid, and restricted videos.

\textbf{Detecting video subtitles.} In order to guarantee the quality of our corpora annotations to the greatest extent, we  implement a subtitle detection process for videos, filtering out those that feature manually uploaded subtitles. The rest of the videos that lack subtitles are relegated to function as unsupervised data sources, utilizing solely their audio components.

\textbf{Downloading audio and subtitles.} We download the audio tracks of videos and their corresponding manually uploaded subtitles through the YouTube download engine\footnote{https://github.com/yt-dlp/yt-dlp} as the primary data source for  MSR-86K. Additionally, we download the audio from some videos without subtitles to serve as the data source for unsupervised pre-training. Each audio file is converted into a single-channel wav format, sampled at a 16 kHz rate.

\subsection{ASR corpus construction}
\textbf{Text normalization.} Video subtitles contain several non-semantic symbols. To streamline further processing, we need to normalize the text. This involves transforming the case, removing punctuation and emojis, converting numbers, and eliminating special symbols associated with specific languages.

\textbf{Forced alignment.} Even though video subtitles come with timestamps, we often notice a lot of them aren't accurate, thus necessitating a re-alignment of the audio with the subtitles. Thanks to the work of predecessors, we use a pre-trained ASR model based on the connectionist temporal classification (CTC) \cite{kurzinger2020ctc} criterion for alignment, and take the median of the alignment scores as the cutoff for filtering.

\begin{figure}[h]

  \centering
 \includegraphics[width=1.0\linewidth]{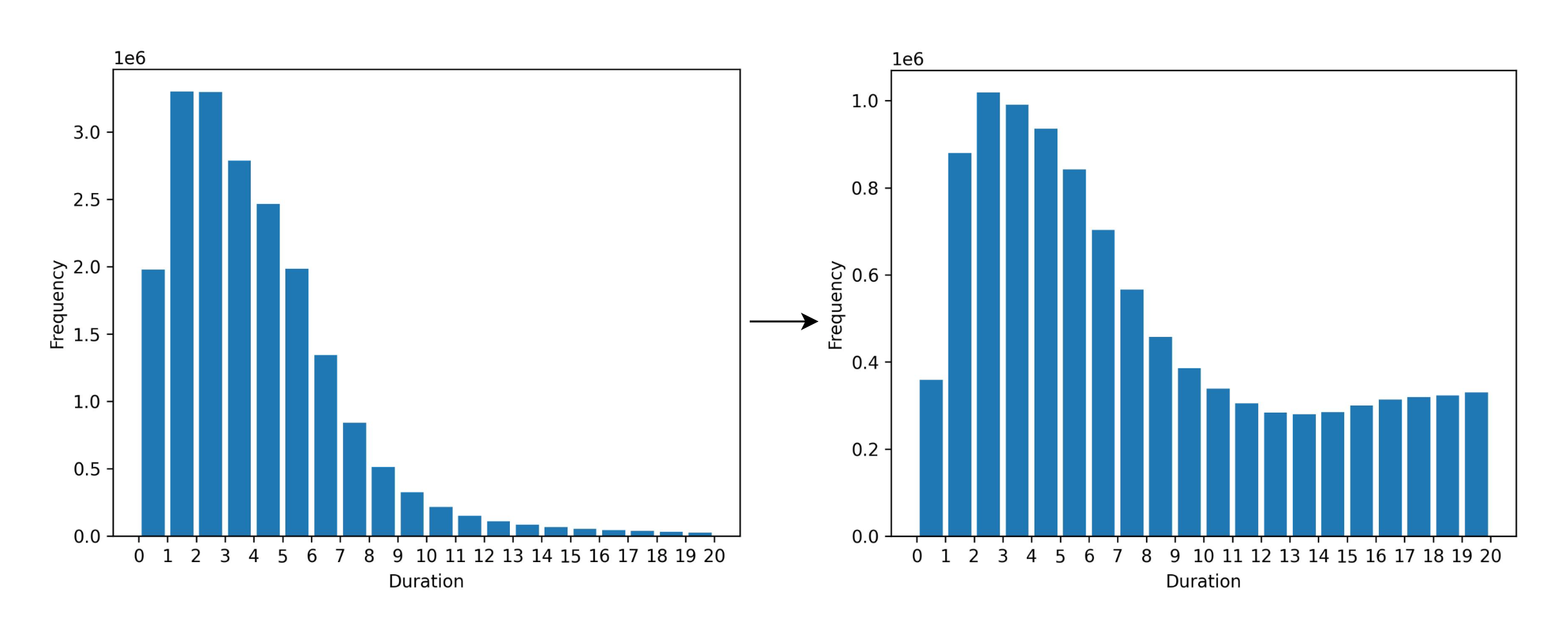}
  \caption{The description of duration balance.}
  \label{fig:speech_production}
\end{figure}

\textbf{Duration balance.}  Due to memory constraints, subtitles are usually segmented for forced alignment. However, each segment does not necessarily correspond to the exact endpoint of the speaker's utterance, resulting in a relatively short distribution of audio duration. To balance the audio duration and ensure the integrity of the speech content as much as possible, we conducted voice activity detection (VAD) based on the output of the CTC model, and limited the maximum duration to 20 seconds. Figure 2 shows the duration statistics before and after VAD.

\textbf{LID filter.}  After reviewing the outcomes of forced alignment, we noticed that there were still some inaccuracies. The most common issues included mismatched languages between the audio and the subtitles, subtitles that were categorized as descriptive captions, and audio that was either purely music or completely silent. Consequently, we develop a  language identification (LID) model that effectively filters out sentences where discrepancies exist between the audio and the subtitles, significantly improving the quality of the data.

\textbf{ASR filter.}  To further improve data quality, we train an ASR model using both existing open-source data and the data filtered by the LID model. This ASR model is used to decode the data processed by the LID filter and calculate the word error rate (WER). By filtering out segments with higher WER, we ensure greater accuracy in our dataset annotations.

\begin{table}[htbp]
  \centering
  \caption{Duration of different languages in the MSR-86K corpus and the results of the Monolingual ASR on the dev set.}
    \begin{tabular}{lrrc}
    \toprule
    \multicolumn{1}{c}{\multirow{2}[4]{*}{\textbf{Language}}} & \multicolumn{2}{c}{\textbf{Duration (hrs)}} & \multicolumn{1}{p{5em}}{\textbf{Monolingual  }} \\
\cmidrule{2-4}          & \multicolumn{1}{l}{~~~~train} & \multicolumn{1}{l}{~~~~dev} & \multicolumn{1}{l}{WER/CER (\%)} \\
    \midrule
    Spanish & 13976.84  & 18.63 & 6.36  \\
    Korean & 10338.66  & 18.56 & 4.79  \\
    English & 9795.46  & 17.42 & 5.61  \\
    French & 8316.70  & 15.84 & 8.43  \\
    German & 6862.00  & 14.38 & 6.39  \\
    Hindi & 5986.50  & 11.62 & 9.90  \\
    Vietnamese & 5957.47  & 11.54 & 3.51  \\
    Italian & 5691.28  & 11.40  & 5.18  \\
    Dutch & 4138.50  & 9.67  & 5.99  \\
    Portuguese & 3737.54  & 9.62  & 7.43  \\
    Thai  & 3674.70  & 9.47  & 4.17  \\
    Russian & 3188.52  & 9.35  & 8.90  \\
    Indonesian & 1982.87  & 9.12  & 6.75  \\
    Japanese & 1779.03  & 8.54  & 3.44  \\
    Arabic & 873.84  & 4.95  & 9.40  \\
    \midrule
    Total/Avg & 86299.91 & 180.11 & 6.42  \\
    \bottomrule
    \end{tabular}%
  \label{tab:addlabel}%
\end{table}%

\textbf{Data split.} Based on the forced alignment scores, LID scores, and WER, we select a portion of the data with the highest quality to serve as the development set, while the remaining data is allocated for the training set. The distribution of durations across different languages is detailed in Table 2. For the test set, we use the  test portion of the Common Voice corpus, which has undergone stringent manual verification to ensure the high quality required for multilingual ASR testing.

\subsection{Unsupervised corpus construction}
For audio without subtitles, we employ a sound event detection model to filter out music and noise, and segment the audio at points of silence into clips shorter than 30 seconds. Ultimately, we obtain a total of 200k hours of unsupervised data.

\section{Experiments and Discussions}
In this section, we first introduce the evaluation of the MSR-86K corpus, assessing the overall quality of the corpus. Secondly, we describe how to use unsupervised data for pre-training, and then fine-tune with MSR-86K and other open-source data to obtain a non-autoregressive multilingual speech recognition model that outperforms the whisper large model.

\subsection{Data evaluation}
To evaluate the quality of the MSR-86K corpus, we trained a monolingual model using the training set for each language. Then, we performed Beam Search decoding on the MSR-86K development set and calculated the word error rate  and character error rate (CER). Our evaluation model utilizes the Transformer-CTC architecture, in which  $d^{model}$=768, ${d^{ffn}}$=3072, ${d^{head}}$=12 , $num\_layers$=12. In addition, a convolutional front-end was used to sub-sample the acoustic features by a factor of 6. Moreover, each language is equipped with its own respective vocabulary, which employs a byte-level  byte-pair encoding (BPE) model with a vocabulary size of 2000.

As shown in Table 2, the monolingual ASR models for 15 languages all achieved a WER or CER  below 10$\%$  on their respective development sets, with some languages reaching below 5$\%$ , and an average error rate of 6.42$\%$  across all languages. Considering that our evaluation model does not employ state-of-the-art ASR models and given the spontaneous nature of YouTube audio, an overall error rate of 6.42$\%$ meets our expectations, indicating that the data quality has reached a relatively ideal level. Therefore, in practice, the development set of MSR-86K can serve as a multilingual test set of the YouTube domain for other open-source  corpora.

\begin{figure}[th]
  \centering
  \includegraphics[width=1.0\linewidth]{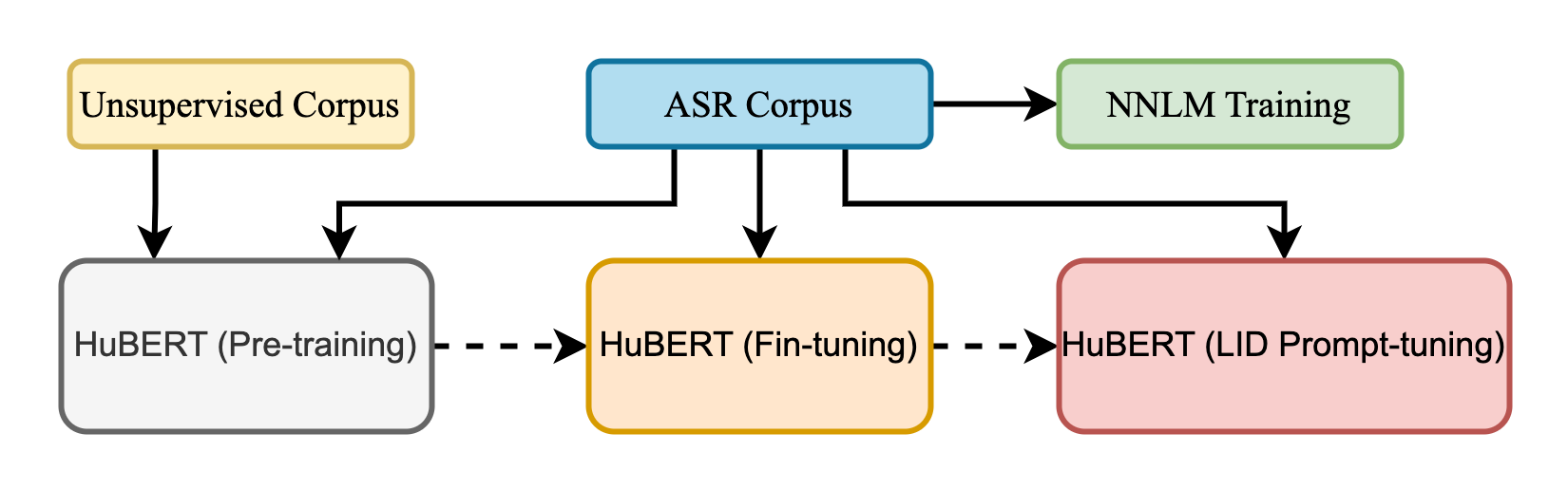}
  \caption{The workflow of our multilingual ASR model training.}
\end{figure}

\subsection{Multilingual ASR construction}
Whisper is an excellent multilingual model that performs well across mainstream languages around the world. However, Whisper has not made its training data public, making it difficult for researchers to replicate its results. Our MSR-86K corpus effectively bridges this gap and can facilitate researchers' studies on large-scale multilingual speech recognition. Additionally, the best-performing model of Whisper has a high parameter count of up to 1.55 billion, which results in slower inference speed and also requires more memory and computational resources. In this section, we explain how to leverage easily accessible unsupervised data for pre-training, and then fine-tune with the MSR-86K and other existing multilingual open-source corpora to develop a multilingual ASR model that has a smaller parameter size, faster speed, and performance that is comparable to or even surpasses that of Whisper. The workflow of our multilingual ASR model training is illustrated in Figure 3.

\textbf{Data preparation.} Whisper (v2) was trained using 680k hours of annotated data, while Whisper larger-v3 has reached a scale of 5 million hours, which is daunting for the average researcher. As illustrated in Table 3, we employed our contributed MSR-86K and various other open-source multilingual corpora for our transcribed data. In addition, to reduce the model's dependency on transcribed data, we explored unsupervised pre-training methods. By leveraging the data listed in Table 3 and incorporating the unsupervised data detailed in Section 2.3, we amassed a comprehensive corpus of 400k hours.

\begin{table}[htbp]
  \centering
  \caption{Summary of multilingual open-source corpora for our experiments.}
\resizebox{70mm}{49mm} {
    \begin{tabular}{lrr}
    \toprule
    \textbf{Language} & \multicolumn{1}{l}{\textbf{Corpus}} & \multicolumn{1}{l}{\textbf{Total Hours}} \\
    \midrule
    English & \multicolumn{1}{p{15.415em}}{Librispeech\cite{panayotov2015librispeech}, mTEDx\cite{salesky2021multilingual},  GigaSpeech\cite{chen2021gigaspeech}, MLS\cite{pratap2020mls}, The Peoples's Speech\cite{galvez2021people}, CommonVoice\cite{ardila2019common}, MSR-86K} & 78180.96 \\
   \midrule  Chinese & \multicolumn{1}{p{15.415em}}{AISHELL-1\cite{bu2017aishell}, AISHELL-2\cite{du2018aishell}, Thchs30\cite{wang2015thchs}, Aidatatang\_200zh\cite{aidata}, \newline{}Primewords\cite{choi2018pansori}, TAL\_ASR\cite{talasr}, TAL\_CSASR\cite{li2022talcs},  WenetSpeech\cite{zhang2022wenetspeech}, CommonVoice} & 12227.58 \\
    \midrule Spanish & \multicolumn{1}{p{15.415em}}{mTEDx, CommonVoice, MLS, \newline{}MSR-86K} & 15507.07 \\
    Korean & \multicolumn{1}{p{15.415em}}{CommonVoice, Zeroth\_Korean\cite{zero}, MSR-86K} & 10418.17 \\
    French & \multicolumn{1}{p{15.415em}}{CommonVoice, MLS, MSR-86K} & 10125.3 \\
    German & \multicolumn{1}{p{15.415em}}{Bundestag\cite{wirth2023asr}, MLS, CommonVoice, \newline{}MSR-86K} & 10278.88 \\
    Hindi & \multicolumn{1}{p{15.415em}}{CommonVoice, MSR-86K} & 5994.73 \\
    Vietnamese & \multicolumn{1}{p{15.415em}}{CommonVoice, MSR-86K} & 5960.94 \\
    Italian & \multicolumn{1}{p{15.415em}}{CommonVoice, MSR-86K, MLS} & 6183.40 \\
    Dutch & \multicolumn{1}{p{15.415em}}{CommonVoice, MSR-86K, MLS} & 5750.03 \\
    Portuguese & \multicolumn{1}{p{15.415em}}{CommonVoice, MSR-86K, MLS, \newline{}mTEDx} & 4074.15 \\
    Thai  & \multicolumn{1}{p{15.415em}}{CommonVoice, MSR-86K} & 3726.11 \\
    Russian & \multicolumn{1}{p{15.415em}}{CommonVoice, Open\_STT\cite{openstt}, Golos\cite{karpov2021golos}, MSR-86K} & 8149.94 \\
    Indonesian & \multicolumn{1}{p{15.415em}}{CommonVoice, MSR-86K} & 1994.47 \\
    Japanese & \multicolumn{1}{p{15.415em}}{CommonVoice, MSR-86K, JTubeSpeech\cite{takamichi2021jtubespeech}, Reazonspeech\cite{fujimotoreazonspeech}} & 21774.56 \\
    Arabic & \multicolumn{1}{p{15.415em}}{CommonVoice, MSR-86K, MGB2\cite{ali2016mgb}, QASR\cite{mubarak2021qasr}} & 4162.44 \\
    \midrule
    Total &       & 204508.73 \\
    \bottomrule
    \end{tabular}}%
  \label{tab:addlabel}%
\end{table}%

\begin{table*}[htbp]
  \centering
  \caption{Compare our multilingual ASR model and Whisper model in terms of WER/CER ($\%$ ) on the open-source test set.}
\resizebox{170mm}{37mm} {
    \begin{tabular}{lrcccccccc}
    \toprule
    \multicolumn{1}{c}{\multirow{2}[4]{*}{\textbf{Language}}} & \multicolumn{1}{c}{\multirow{2}[4]{*}{\textbf{Test set}}} & \multicolumn{5}{c}{\textbf{Whisper}}  &       & \multicolumn{2}{c}{\textbf{Ours}} \\
\cmidrule{3-7}\cmidrule{9-10}          &       & \multicolumn{2}{p{20.415em}}{~~~~~~~~~~~~~~~~~~~~~~~~~~~~~Medium (769M) } &       & \multicolumn{2}{p{20.915em}}{~~~~~~~~~~~~~~~~~~~~~~~~~~~~~Larger v2 (1.55B) } &       & \multicolumn{2}{p{21.33em}}{~~~~~~~~~~~~~~~~~~~~~~~~~~~~~HuBERT-CTC (362M)} \\
\cmidrule{1-4}\cmidrule{6-10}          &       & \multicolumn{1}{p{10.25em}}{~~~~~~~~~~~without LID} & \multicolumn{1}{p{10.165em}}{~~~~~~~~~~~~with LID} &       & \multicolumn{1}{p{10.75em}}{~~~~~~~~~~~without LID} & \multicolumn{1}{p{10.165em}}{~~~~~~~~~~~~with LID} &       & \multicolumn{1}{p{10.165em}}{~~~~~~~~~~~without LID} & \multicolumn{1}{p{11.165em}}{~~~~~~~~~~~~~~with LID} \\
\cmidrule{1-7}\cmidrule{9-10}    \multirow{6}[1]{*}{English} & \multicolumn{1}{l}{~~~~~~~Librispeech test-clean} & 3.5   & 2.8   &       & 3.4   & 2.6   &       & 2.5   & 2.3  \\
          & \multicolumn{1}{l}{~~~~~~~Librispeech test-other} & 6.5   & 6.2   &       & 5.3  & 4.9   &       & 5.2   & 4.9  \\
          & \multicolumn{1}{l}{~~~~~~~Gigaspeech} & 13.6  & 12.7  &       & 11.8  & 10.6  &       & 11.8  & 10.5  \\
          & \multicolumn{1}{l}{~~~~~~~MLS} & 8.1  & 7.0   &       & 7.3   & 6.8   &       & 6.6   & 6.4  \\
          & \multicolumn{1}{l}{~~~~~~~TEDLIUM2} & 4.6   & 4.5   &       & 4.5   & 4.4   &       & 4.3   & 4.2  \\
          & \multicolumn{1}{l}{~~~~~~~CommonVoice} & 22.4  & 13.4  &       & 17.6  & 10.8  &       & 11.2  & 10.7  \\
  \midrule
    \multirow{6}[1]{*}{Chinese} & \multicolumn{1}{l}{~~~~~~~AISHELL-1} & 7.1   & 7.3   &       & 6.2   & 6.2   &       & 3.2   & 2.8  \\
          & \multicolumn{1}{l}{~~~~~~~AISHELL-2 test-mic} & 6.8   & 6.1   &       & 5.3   & 5.3   &       & 4.8   & 4.3  \\
          & \multicolumn{1}{l}{~~~~~~~THCHS-30} & 8.6   & 8.6   &       & 6.8   & 6.8   &       & 6.2   & 6.0  \\
          & \multicolumn{1}{l}{~~~~~~~WenetSpeech dev} & 11.7  & 11.5  &       & 10.4  & 10.4  &       & 7.1   & 6.5  \\
          & \multicolumn{1}{l}{~~~~~~~WenetSpeech test-meeting} & 22.0  & 22.5  &       & 22.1  & 21.6  &       & 12.7  & 10.1  \\
          & \multicolumn{1}{l}{~~~~~~~CommonVoice} & 16.1  & 16.0  &       & 15.7  & 14.2  &       & 10.2  & 9.8  \\
    \midrule
    Spanish & \multicolumn{1}{l}{\multirow{14}[2]{*}{~~~~~~~CommonVoice}} & 9.9  & 7.8   &       & 6.5   & 6.3   &       & 6.3   & 5.8  \\
    Korean &       & 7.9   & 7.9   &       & 6.2   & 6.2   &       & 6.0   & 5.8  \\
    French &       & 17.9  & 14.7  &       & 12.9  & 11.6  &       & 10.4  & 10.1  \\
    German &       & 10.1  & 8.3   &       & 7.6   & 6.5   &       & 6.4   & 6.2  \\
    Hindi &       & 62.0  & 46.4  &       & 53.0  & 37.0  &       & 19.6  & 17.1  \\
    Vietnamese &       & 26.7  & 24.8  &       & 23.2  & 21.3  &       & 12.3  & 10.0  \\
    Italian &       & 12.2  & 9.3   &       & 8.9   & 7.6   &       & 7.7   & 7.2  \\
    Dutch &       & 10.5  & 7.8   &       & 7.6   & 5.8   &       & 5.9   & 5.6  \\
    Portuguese &       & 11.9  & 8.8   &       & 8.9  & 7.2   &       & 6.9   & 6.5  \\
    Thai  &       & 14.1  & 12.0  &       & 10.9  & 10.3  &       & 3.2   & 3.0  \\
    Russian &       & 12.0  & 10.6  &       & 8.6   & 8.1   &       & 8.6   & 7.5  \\
    Indonesian &       & 13.8  & 13.0  &       & 11.5  & 8.7   &       & 8.1   & 8.0  \\
    Japanese &       & 14.5  & 11.4  &       & 12.1  & 9.7  &       & 10.5  & 9.6  \\
    Arabic &       & 31.7  & 30.7  &       & 21.8  & 21.2  &       & 19.8  & 15.5  \\
    \midrule
    Avg   &       & 14.9  & 12.8  &       & 12.2  & 10.5  &       & 8.4   & 7.6  \\
    \bottomrule
    \end{tabular}}%
  \label{tab:addlabel}%
\end{table*}%

\begin{table}[htbp]
  \centering
  \caption{Comparison of WER/CER($\%$) on the MSR-86K development set for Whisper and our system with LID provided in advance.}
\resizebox{80mm}{28mm} {
    \begin{tabular}{lcccc}
    \toprule
    \multicolumn{1}{c}{\multirow{2}[4]{*}{\textbf{Language}}} & \multicolumn{2}{c}{\textbf{Whisper}} &       & \textbf{Ours} \\
\cmidrule{2-3}\cmidrule{5-5}          & Medium (769M)  & Larger v2 (1.55B)  &       & HuBERT-CTC (362M) \\
    \midrule
    English & 4.9   & 4.4   &       & 4.0  \\
    Spanish & 7.4   & 7.1   &       & 5.6  \\
    Korean & 9.6   & 7.2   &       & 4.5  \\
    French & 9.8   & 9.4   &       & 7.0  \\
    German & 6.3   & 5.6   &       & 5.0  \\
    Hindi & 69.0  & 47.4  &       & 7.5  \\
    Vietnamese & 28.8  & 19.2  &       & 3.1  \\
    Italian & 6.2   & 5.8   &       & 4.1  \\
    Dutch & 8.5   & 7.3   &       & 5.1  \\
    Portuguese & 8.7   & 7.8   &       & 6.1  \\
    Thai  & 27.9  & 21.7  &       & 3.0  \\
    Russian & 9.7   & 9.7   &       & 7.3  \\
    Indonesian & 13.7  & 11.0  &       & 5.0  \\
    Japanese & 7.6   & 6.9   &       & 3.4  \\
    Arabic & 67.4  & 44.5  &       & 6.3  \\
    \midrule
    Avg   & 19.0  & 14.3  &       & 5.1  \\
    \bottomrule
    \end{tabular}}%
    \vspace{-0.6cm}
  \label{tab:addlabel}%
\end{table}%

\textbf{Pre-training.} We first conducted unsupervised pre-training with the prepared data. Given the superior performance of HuBERT\cite{hsu2021hubert}, we chose it as the criterion for unsupervised pre-training.  We used a Transformer encoder similar to the one described in Section 3.1 as the acoustic encoder, where $d^{model}$=1024, ${d^{ffn}}$=4096, ${d^{head}}$=16, $num\_layers$=24. 

\textbf{Fine-tuning.} Next, we fine-tuned the pre-trained HuBERT model using the dataset presented in Table 3, with CTC as the training criterion. Similar to Whisper, our vocabulary is shared across all languages. We trained a byte-level BPE model with a vocabulary size of 10,000 using the texts from the corpora presented in Table 3 to establish the lexicon for CTC, to which we added an extra token to signify the blank symbol.

 \textbf{LID Prompt-tuning.}  Multilingual ASR typically encounters two usage scenarios. The first scenario is where the language of the speech to be recognized is not known in advance, necessitating the model to identify it autonomously.  In the second, the language information is provided in advance,  guiding the model to bolster its performance in recognizing the specified language. To enable the CTC model to accommodate both scenarios, we employed the method  proposed in\cite{li2023enhancing}, using language identity (LID) as a prompt to enhance the recognition performance of the target language.

\textbf{NNLM Training. }To further enhance the performance of the HuBERT-CTC multilingual ASR model, we trained a simple LSTM-based language model using the text from the corpora in Table 3, and employed shallow fusion for decoding.

Through the four steps mentioned above, we  obtained a high-performance multilingual ASR model with a total parameter size of 362M, which is substantially smaller than the Whisper larger model,  making it more suitable for  deployment.

\subsection{Multilingual ASR evaluation}
Due to the differences in the scale of training data, our primary benchmark for the multilingual ASR model is the Whisper larger-v2 model. We use the pipeline provided by HuggingFace\footnote{https://huggingface.co/openai/whisper-large-v3} for inference, testing both models where LID is provided in advance and where LID is not provided. Most previous papers have not tested the performance of Whisper without  LID, and we believe that the results without LID are also meaningful. The recognition results for all languages were subjected to text normalization prior to the calculation of WER or CER. It is important to note that for Chinese, converting from traditional to simplified characters is necessary for calculation accuracy.

As shown in Table 4, our multilingual ASR model outperforms the Whisper medium and larger-v2 models across all languages, regardless of whether the LID is provided in advance or not, and was trained with less transcribed data.  It's worth mentioning that Whisper's performance significantly declines on the Common Voice English test set when the LID is not specified beforehand.  This performance dip can be largely ascribed to erroneous LID predictions, which exacerbate the inherent error propagation found in autoregressive models, culminating in less-than-ideal outcomes. On the other hand, our model demonstrates robustness and maintains stable performance, unaffected by the presence or absence of LID information.  The results in Table 5 once again demonstrate that our model surpasses Whisper on the MSR-86K development set, which is indicative of the advanced nature of our algorithms.

\section{Conclusions}
In this paper, we introduce the MSR-86K corpus, an evolving, multilingual corpus with 86,300 hours of transcribed audio for speech recognition research. We believe that such a large-scale  corpus will propel the research in multilingual speech algorithms.    We also hope that more researchers will contribute to open-source data and work together to advance the development of the intelligent speech field. Additionally, we explain  how to effectively leverage readily available unsupervised data, MSR-86K, and other open-source corpora to train a robust ASR model that is competitive with Whisper in terms of performance but smaller in size and faster, allowing everyone to use open-source data to train their own multilingual ASR model.  
\bibliographystyle{IEEEtran}

\bibliography{mybib}


\end{document}